\documentstyle[12pt]{article}
\begin{document}

\noindent Stockholm\\
USITP 04-4\\
June 2004\\

\vspace{1cm}

\begin{center}

{\Large MUBS, POLYTOPES, AND FINITE GEOMETRIES}\footnote{Talk at the 
V\"axj\"o conference on Foundations of Probability and Physics, 
June 2004.} 

\vspace{1cm}

{\large Ingemar Bengtsson}\footnote{Email address: ingemar@physto.se. 
Supported by VR.}

\

{\sl Stockholm University, Alba Nova\\
Fysikum\\
S-106 91 Stockholm, Sweden}

\vspace{8mm}

{\bf Abstract}

\end{center}

\vspace{5mm}

\noindent A complete set of $N+1$ mutually unbiased bases (MUBs) 
exists in Hilbert spaces of dimension $N = p^k$, where $p$ is a 
prime number. They mesh naturally with finite affine planes of 
order $N$, that exist when $N = p^k$. The existence of MUBs for 
other values of $N$ is an open question, and the same is true 
for finite affine planes. I explore the question whether the 
existence of complete sets of MUBs is directly related to the 
existence of finite affine planes. Both questions can be shown 
to be geometrical questions about a convex polytope, but not in 
any obvious way the same question. 

\newpage

{\bf 1. Introduction}

\vspace{5mm}

\noindent The acronym MUBs stands for ``mutually unbiased bases''. 
By definition, two orthonormal bases ${|e_i\rangle }$ and ${|f_i\rangle }$ 
are mutually unbiased if it is true that 

\begin{equation} |\langle e_i|f_j\rangle |^2 = \mbox{constant} = 
\frac{1}{N} \label{1} \end{equation}

\noindent for every choice of $i, j$. Here $N$ is the dimension of the 
Hilbert space, and the second equality gives the value that the constant 
has to take. Equation (\ref{1}) is reminiscent of the relation between 
the eigenbases of position and momentum---if we have complete knowledge of 
one observable, we know nothing about the other. Bases that are mutually 
unbiased underlie notions such as complementarity. Beginning with Weyl 
\cite{Weyl}, they have been studied for a long time 
\cite{Schwinger,Kraus}. I hope they are 
of interest also in a conference on the foundations of probability and 
physics. After all, complementarity is one of the most striking features 
of the best probability theory that physics has to offer, and MUBs are 
part of the mathematics of complementarity. The importance of MUBs for 
the foundations of quantum mechanics has been stressed, especially, 
by Brukner and Zeilinger \cite{Anton}.

The name MUB is due to Wootters, who pointed out---quite some time 
ago \cite{Wootters1}---that if a Hilbert space of dimension $N$ contains 
$N+1$ bases such that every pair of bases is mutually unbiased, then 
there exists a finite dimensional analogue of the 
Wigner function, where the role of the classical phase space is played 
by a finite affine plane of order $N$. The use of finite affine planes 
in quantum mechanics can be traced back---in the qubit case---at least 
to Feynman \cite{Feynman}; for the most complete treatment so far, see 
Gibbons et al. \cite{Wootters3}. Wootters' original motivation 
had to with quantum state tomography. There are other motivations too: 
Both MUBs and finite affine planes play a role in various quantum 
communication protocols \cite{Fivel} and quantum cryptography schemes 
\cite{Peres,Karlsson}---most spectacularly in a story \cite{Aharonov} with 
a happy end \cite{Schulz}, known as the Mean King's Problem 
\cite{Englert}---but my suspicion is that the best is yet to come. 
Finite geometries are used by people who are handling, encrypting 
and deciphering information classically. With 
quantum information theory in ascendancy it is easy to imagine that, in 
the long run, the appearance of finite affine planes in quantum 
mechanics will enable us to perform a wide range of interesting 
tricks. A good omen is that some of 
Wootters' key results were rediscovered, and placed in the context of 
discrete mathematics, by Zauner in his Ph D thesis \cite{Zauner}.

It is known that a Hilbert space of dimension $N$ can contain at most 
$N+1$ MUBs, so we refer to this many MUBs as a complete set. For $N = p$, 
where $p$ is a prime number, a complete set of MUBs was given by 
Ivanovi\'c \cite{Ivanovic}, and for $N = p^k$ by Wootters and Fields 
\cite{Wootters2}.
A quarter of a century later, the question whether 
a complete set of MUBs exists for other values of $N$ remains open. Since I 
have not yet explained what finite affine planes are, or what they have 
to do with MUBs, it may look like an odd fact to mention, but it happens 
that finite affine planes of order $N$ exist if $N = p^k$, they 
do not exist when $N = 6, 10, 14, \dots $ (an infinite set of values are 
excluded), and it is an open question whether they exist when $N = 12, 15, 
20, \dots $ (an infinite set of values are open). This question goes 
back to Euler, so after all these years it appears justified to say that 
it is a hard one. Anyway, for discussion we have two 
questions to choose from:

\smallskip
{\bf I}: How many MUBs exist in a given Hilbert space?
 
{\bf II}: Given a complete set of MUBs, what can we do with them?
\smallskip

\noindent I choose the first question, since I have nothing to say about 
the second. I do not have an answer to the first either, and in fact I am 
not sure what kind of question it is. We can discuss it as a close packing 
problem in complex projective space \cite{Conway}, or we can try to 
phrase it as a question of finding $N$ complex Hadamard matrices related 
in a particular way---which would tie in with what I talked about in 
last year's V\"axj\"o meeting \cite{IB}. In fact the set of $N\times N$ 
complex Hadamard matrices has quite special properties when $N$ is prime, 
or a power of a prime \cite{Wojtek}. 
However, as it stands this observation explains nothing---except why I give 
this talk this year---so I choose to explore the obvious question whether 
the existence of a complete set of MUBs is related to the existence of affine 
planes. The dream is to use one of these problems to solve the other!

I should perhaps say, at the start, that the only conclusion I will arrive at 
is that an affine plane of order $N$ exists if and only if it is possible to 
inscribe a regular simplex with $N^2$ corners in the polytope defined by 
the MUBs (assuming that the latter exist). I would like to think that this 
sheds some light on the question, but I am not so sure. 

\newpage
 
\vspace{1cm}

{\bf 2. The Complementarity Polytope}

\vspace{5mm}

\noindent I will work with density matrices rather than state vectors. An ON basis 
in Hilbert space then becomes a set of $N$ projectors $P_i = |e_i\rangle 
\langle e_i|$. These operators sit in the $N^2 - 1$ dimensional Euclidean space 
of Hermitian matrices of unit trace, with the distance squared between two 
matrices given by 

\begin{equation} D^2(A,B) = \frac{1}{2}\mbox{Tr}(A-B)^2 \ . \end{equation}

\noindent The factor $1/2$ ensures that the projectors in the basis form the 
$N$ corners of a regular $N-1$ dimensional simplex with edges of unit length. 
Since we will encounter several different kinds of simplices, of different 
dimensionalities, I will refer to these simplices as B-simplices. 

We would like to regard our space as a vector space. We can do this by placing 
the origin at the matrix

\begin{equation} \rho_* = \frac{1}{N} \ . \end{equation}

\noindent The scalar product of two unit trace matrices $A$ and $B$ is then given 
by 

\begin{equation} (A,B) = \frac{1}{4}\left[ D^2(A+B,\rho_*) - 
D^2(A-B,\rho_*)\right] = \frac{1}{2}\left[ \mbox{Tr}AB - \frac{1}{N}\right] 
\ . \end{equation}

\noindent The condition for orthogonality of two matrices becomes $\mbox{Tr}AB = 
1/N$, so in particular a pair of projectors belonging to two different MUBs are 
orthogonal.

We observe that a complete set of MUBs defines a convex polytope in our 
$N^2-1 = (N-1)(N+1)$ dimensional space, which can be described very simply: 
We split our space into $N+1$ perpendicular $N-1$ dimensional subspaces, 
and in each subspace we place a regular simplex with $N$ corners, centered 
at the origin. Taking the convex cover of all the $N^2 + N$ corners we obtain 
a convex polytope that we will call the Complementarity 
Polytope. In terms of matrices, letting the indices $i, j$ run from $1$ to 
$N$ and the indices $I, J$ run from $0$ to $N$, the corners define 
$N(N+1)$ matrices $P_{Ii}$ such that 

\begin{equation} \mbox{Tr}P_{Ii} = \mbox{Tr}P^2_{Ii} = 1 \label{5} \end{equation}

\begin{equation} \mbox{Tr}P_{Ii}P_{Ij} = 0 \ , \hspace{6mm} \small{i \neq j} 
\label{6} \end{equation}

\begin{equation} \mbox{Tr}P_{Ii}P_{Jj} = \frac{1}{N} \ , \hspace{5mm} 
\small{I \neq J} \ . \label{7} \end{equation}

\noindent Equation (\ref{5}) ensures that all corners sit on a sphere of radius 
$\sqrt{(N-1)/2N}$ and centered at $\rho_*$, equation (\ref{6}) ensures that they 
form one regular B-simplex for each value of $I$, and equation (\ref{7}) ensures 
that the B-simplices sit in $N+1$ perpendicular subspaces. Sometimes we will 
find it convenient to use a collective index for the corners, i.e. we will 
refer to $P_{\omega}$, where $\omega $ runs between $1$ and $N^2 + N$.  

Reflecting on the definition, we see that it is easy to form a 
Complementarity Polytope, regardless of the value 
of $N$. What is not easy is to ensure that its corners are matrices with 
non-negative spectrum. If they are not, they are not density matrices. The 
question of the existence of a complete set of MUBs can now be reformulated: 
Given a Complementarity Polytope, and given the convex body of density 
matrices, is it possible to arrange them in such a way that the former 
is a subset of the latter? Evidently this is not going to be a walkover---the 
corners of the Complementarity Polytope lie on the outsphere of the body 
of density matrices, and the set of pure quantum states there form a $2(N-1)$ 
dimensional subset of this $N^2 - 2$ dimensional sphere. When $N > 2$ this 
means that all the corners of the Complementarity Polytope must be fitted 
into a small subset of its outsphere. But we can study the Complementarity 
Polytope regardless of the existence of a complete set of MUBs, since---as 
a polytope---it certainly exists in all $N^2 - 1$ dimensional vector spaces. 
This is exactly what we will do, until close to the end of the talk.

The Complementarity Polytope 
can be visualized easily, at least as long as we are not too demanding 
about how much we want to see. When $N = 2$ it is an octahedron. More 
schematically, we can think of the octahedron as consisting of three 
orthogonal one dimensional B-simplices, i.e. as 
$| \hspace{-2.7mm} 
{\ }^{\bullet}_{\bullet} \hspace {1mm} | \hspace{-2.7mm} 
{\ }^{\bullet}_{\bullet} \hspace {1mm} | \hspace{-2.7mm} 
{\ }^{\bullet}_{\bullet} \hspace {1mm}$. When $N = 3$ 
we are dealing with four totally orthogonal triangles, 
$\triangle \hspace{-5mm} 
{\ }_{\bullet} \hspace{-1.9mm} {\ }^{\bullet} \hspace{-2mm} 
{\ }_{\bullet} \hspace{1.3mm} \triangle \hspace{-5mm} 
{\ }_{\bullet} \hspace{-1.9mm} {\ }^{\bullet} \hspace{-2mm} 
{\ }_{\bullet} \hspace{1.3mm} \triangle \hspace{-5mm} 
{\ }_{\bullet} \hspace{-1.9mm} {\ }^{\bullet} \hspace{-2mm} 
{\ }_{\bullet} \hspace{1.3mm} \triangle \hspace{-5mm} 
{\ }_{\bullet} \hspace{-1.9mm} {\ }^{\bullet} \hspace{-2mm} 
{\ }_{\bullet} \hspace{1.3mm} $, when $N = 4$ we have five totally 
orthogonal tetrahedra, 
$\Box \hspace{-5mm} {\ }^{\bullet}_{\bullet} \hspace{-2.2mm} 
\times \hspace{-3.5mm} 
{\ }^{\bullet}_{\bullet} \hspace{3mm} \Box \hspace{-5mm} {\ }^{\bullet}_{\bullet} 
\hspace{-2.2mm} \times \hspace{-3.5mm} 
{\ }^{\bullet}_{\bullet} \hspace{3mm} \Box \hspace{-5mm} {\ }^{\bullet}_{\bullet} 
\hspace{-2.2mm} \times \hspace{-3.5mm} 
{\ }^{\bullet}_{\bullet} \hspace{3mm} \Box \hspace{-5mm} {\ }^{\bullet}_{\bullet} 
\hspace{-2.2mm} \times \hspace{-3.5mm} 
{\ }^{\bullet}_{\bullet} \hspace{3mm} \Box \hspace{-5mm} {\ }^{\bullet}_{\bullet} 
\hspace{-2.2mm} \times \hspace{-3.5mm} 
{\ }^{\bullet}_{\bullet} \hspace{1mm} $, and so on for any $N$. We will be 
interested in the face structure of our polytope, and more especially in 
the following features: When 
$N > 2$ all edges are extremal. There are $N^{N+1}$ faces formed by taking 
the convex cover of one corner from each simplex (in the $N=3$ case this is 
$\triangle \hspace{-4mm} 
{\ }^\bullet \hspace{2.3mm} \triangle \hspace{-4mm} {\ }^\bullet \hspace{2.3mm}  
\triangle \hspace{-4mm} {\ }^\bullet \hspace{2.3mm} 
\triangle \hspace{-4mm} {\ }^\bullet \hspace{2.5mm}$, and its relatives). 
These faces are themselves $N$ dimensional regular simplices, and I call them 
point faces for a reason that will become clear later on, at least to those of 
you who are familiar with Wootters' phase points \cite{Wootters1, Wootters3}. 
To each point face we will associate a point face operator 

\begin{equation} A_{\alpha} = \sum_{point \ face}P_{\omega} - N\rho_*\ , 
\label{8} \end{equation}

\noindent where the index $\alpha $ labels the point faces and the sum runs 
over all the corners in the point face simplex. A more understandable form of 
the same equation is 

\begin{equation} A_{\alpha} - \rho_* = \sum_{point \ face}(P_{\omega} - \rho_*) 
\ . \label{9} \end{equation}

\noindent It is immediate that 

\begin{equation} \mbox{Tr}A_{\alpha} = 1 \hspace{12mm} \mbox{Tr}A^2_{\alpha}
= N \ . \label{10} \end{equation}

\noindent In our vector space the point face operator 
corresponds to a point on the ray from the origin through the center 
of the point face. If continued in the other direction, this ray hits the 
center of a facet, that is the convex cover of $N-1$ corners 
from each of the B-simplices (in the $N=3$ case 
$\triangle \hspace{-5mm} {\ }_{\bullet} \hspace{-0.8mm} 
{\ }_{\bullet} \hspace{2.3mm} \triangle \hspace{-5mm} {\ }_{\bullet} 
\hspace{-0.8mm} {\ }_{\bullet} \hspace{2.3mm} \triangle \hspace{-5mm} 
{\ }_{\bullet} \hspace{-0.8mm} {\ }_{\bullet} \hspace{2.3mm} 
\triangle \hspace{-5mm} {\ }_{\bullet} \hspace{-0.8mm} {\ }_{\bullet} 
\hspace{1mm}$, and its relatives). Hence point faces and facets are in 
one to one correspondence, and they are placed opposite to each 
other on the surface of the polytope. The case $N = 2$ is special since its 
point faces are also facets, and conversely. 

Note that every point face operator defines a set of parallel hyperplanes 
through the equation 

\begin{equation} \mbox{Tr}A\rho = \mbox{constant} \ , \end{equation}

\noindent where $\rho$ is an Hermitian matrix of unit trace. 
When $\rho$ lies in a facet we find that $\mbox{Tr}A\rho = 0$, and when 
$\rho $ lies in a point face we find that $\mbox{Tr}A\rho = 1$. This 
is a strict proof that the faces that we are looking at are extremal. 
It also proves an interesting conjecture by Galv\~ao \cite{Galvao}.   

\newpage

\vspace{1cm}

{\bf 3. Finite affine planes}

\vspace{5mm}

\noindent Now we come to a seemingly odd observation, namely that the number of 
corners of our polytope---$N^2 + N$---is equal to the number of lines in a finite 
affine plane of order $N$. In the best spirit of Euclid, we give the axioms that 
define affine planes. An affine plane is an ordered pair of two sets, 
the first of which consists of elements $a_{\alpha}$, called points, and the 
second of which consists consists of subsets $l_{\omega}$ of the first, 
called lines. Two lines whose intersection is empty are called parallel. 
The following axioms hold:

\

\noindent A1: If $a_{\alpha}$ and $a_{\beta}$ are distinct points, there is 
a unique line $l_{\omega}$ such that $a_{\alpha} \in l_{\omega}$ and 
$a_{\beta} \in l_{\omega}$.

\noindent A2: If $a_{\alpha}$ is a point not contained in the line $l_{\omega}$, 
there is a unique line $l_{\sigma}$ such that $a_{\alpha} \in l_{\sigma}$ 
and $l_{\sigma}\cap l_{\omega} = \emptyset $.  

\noindent A3: There are at least two points on each line, and there are 
at least two lines.

\

\noindent To see how this works, think of an ordinary affine 
plane, and try to think of it as two sets, the set of points and the 
set of lines. Two points determine a unique line, while two lines 
either intersect in a unique point, or else they are parallel and do 
not intersect at all. This is precisely what the axioms say, and it is 
enough to begin to understand how the finite affine planes work.   

We will need only some simple properties. First we define 
a pencil of parallel lines as a maximal set of mutually non-intersecting 
lines. Then we learn, perhaps by looking into some accessible reference 
\cite{Bennett}, that if the number of points is finite, then every line 
contains the same number of points. An affine plane of order $N$ has 
$N$ points on each line, $N^2$ points, and $N^2 + N$ lines. Each point 
is on $N+1$ lines and each pencil of parallel lines contains $N$ lines. 
There are altogether $N+1$ such pencils.

To construct an affine plane of order $N$ we begin by forming a quadratic array 
of $N^2$ points, representing two pencils of parallel lines. We must then 
somehow find $N-1$ additional ways of dividing these points into pencils 
of parallel lines, in such a way that a line from one pencil always shares 
exactly one point with a line from any other pencil. For $N = 2$ it is easy:

\begin{equation} \mbox{1st pencil:} \hspace{5mm} 
\begin{array}{cc} \bullet & \circ \\ 
\bullet & \circ \end{array} \hspace{1cm} 
\begin{array}{cc} \circ & \bullet \\ 
\circ & \bullet \end{array} \hspace{8mm} \mbox{2nd pencil:} \hspace{5mm} 
\begin{array}{cc} \circ & \circ \\ 
\bullet & \bullet \end{array} \hspace{1cm} 
\begin{array}{cc} \bullet & \bullet \\ 
\circ & \circ \end{array} \end{equation}

\begin{equation} \mbox{3d pencil:} \hspace{5mm} 
\begin{array}{cc} \circ & \bullet \\ 
\bullet & \circ \end{array} \hspace{1cm} 
\begin{array}{cc} \bullet & \circ \\ 
\circ & \bullet \end{array} \end{equation}

\noindent For $N = 3$ we get, after slightly more thinking, 

\begin{equation} \mbox{1st pencil:} \hspace{5mm} 
\begin{array}{ccc} \bullet & \circ & \circ \\ 
\bullet & \circ & \circ \\ \bullet & \circ & \circ \end{array} \hspace{1cm}
\begin{array}{ccc} \circ & \bullet & \circ \\ 
\circ & \bullet & \circ \\ \circ & \bullet & \circ \end{array} \hspace{1cm}
\begin{array}{ccc} \circ & \circ & \bullet \\ 
\circ & \circ & \bullet \\ \circ & \circ & \bullet \end{array} \end{equation}

\begin{equation} \mbox{2nd pencil:} \hspace{5mm} 
\begin{array}{ccc} \circ & \circ & \circ \\ 
\circ & \circ & \circ \\ \bullet & \bullet & \bullet \end{array} \hspace{1cm}
\begin{array}{ccc} \circ & \circ & \circ \\ 
\bullet & \bullet & \bullet \\ \circ & \circ & \circ \end{array} \hspace{1cm}
\begin{array}{ccc} \bullet & \bullet & \bullet \\ 
\circ & \circ & \circ \\ \circ & \circ & \circ \end{array} \end{equation}

\begin{equation} \mbox{3d pencil:} \hspace{5mm} 
\begin{array}{ccc} \circ & \circ & \bullet \\ 
\circ & \bullet & \circ \\ \bullet & \circ & \circ \end{array} \hspace{1cm}
\begin{array}{ccc} \bullet & \circ & \circ \\ 
\circ & \circ & \bullet \\ \circ & \bullet & \circ \end{array} \hspace{1cm}
\begin{array}{ccc} \circ & \bullet & \circ \\ 
\bullet & \circ & \circ \\ \circ & \circ & \bullet \end{array} \end{equation}

\begin{equation} \mbox{4th pencil:} \hspace{5mm} 
\begin{array}{ccc} \circ & \bullet & \circ  \\ 
\circ & \circ & \bullet \\ \bullet & \circ & \circ \end{array} \hspace{1cm}
\begin{array}{ccc} \circ & \circ & \bullet \\ 
\bullet & \circ & \circ \\ \circ & \bullet & \circ \end{array} \hspace{1cm}
\begin{array}{ccc} \bullet & \circ & \circ \\ 
\circ & \bullet & \circ \\ \circ & \circ & \bullet \end{array} \end{equation}

\noindent We see that an affine plane is a solution to a combinatorial 
problem.

As $N$ grows it becomes more difficult to 
ensure that lines from different pencils intersect once and once only, 
so it comes as no surprise to learn that it is actually impossible for 
some values of $N$ (such as $N = 6, 10, 14, \dots $). Perhaps it is 
surprising that mathematicians are at a loss to say---even though they have 
had ample time to study the question---whether affine planes exist for certain other 
values of $N$ (such as $N = 12, 15, 20, \dots $). The only values of $N$ 
for which solutions are known are $N = p^k$, where $p$ is an arbitrary prime 
number. In fact for prime powers the existence of finite affine planes is 
guaranteed, because for these values of $N$ affine planes can 
be coordinatized using finite fields, in just the same way that the ordinary 
affine plane can be coordinatized by real numbers. A 
field is a set of ``numbers'' that we can add and multiply, in 
exactly the way we are used to from our experience with rational, real, 
and complex numbers. For any $N$ we can do addition and subtraction modulo $N$, 
but the existence of an inverse to multiplication is non-trivial. It works 
for $N = p$, and---after a step resembling that from real to complex 
numbers---also for $N = p^k$, but there are no further possibilities. 
This does not in itself settle 
the issue of existence of affine planes of order $N$, because there do exist 
affine planes that are unrelated to fields. This happens for instance for 
$N = 9$, and indeed all known examples occur for $N$ being a power of a prime 
\cite{Bennett}. But these examples at least serve to make it conceivable that 
the combinatorics could work out also for, say, $N = 12$.  

\vspace{1cm}

{\bf 4. Painting facets black}

\vspace{5mm}

\noindent Our question is: Can we associate an affine plane of order $N$ with 
the Complementarity Polytope? We can identify the lines with the corners, 
but exactly what collection of $N^2$ parts of the polytope can serve as points?  
It turns out that what we can do is to select a subset of point faces 
and declare that they are to serve as points. Equivalently---because 
of the one to one correspondence between point faces and facets---we can 
select a subset of $N^2$ facets, paint them black, and declare that the 
affine plane consists of the black facets and all the corners. This is 
perhaps the most picturesque way to proceed, and might be the preferred 
way of an $N^2-1$ dimensional being---should he stoop to consider a finite 
plane---but for us it is better to work with the point faces. The combinatorics 
will be the same. 

We tackle it by considering the purely geometrical problem of inscribing 
a regular simplex with $N^2$ corners in the Complementarity Polytope, in such a 
way that each corner of the simplex sits at the center of a point face. 
We call it a D-simplex since it has full dimension. Equivalently, let the 
corners of the D-simplex sit outside the polytope, on 
points represented by the point face operators $A_{\alpha}$. We get a 
regular D-simplex if, restricting the index $\alpha $ to range from $1$ to 
$N^2$, we can choose a set of point face operators so that 

\begin{equation} \mbox{Tr} A_{\alpha}A_{\beta} = N\delta_{\alpha \beta } 
\ . \label{17} \end{equation}

\noindent If so, our selected set of point face operators actually forms 
an orthonormal basis in the $N^2$ dimensional set of Hermitian matrices, 
or in other words in the Lie algebra of $U(N)$. But can we do it? Recalling 
the definition (\ref{8}), which tells us that each point face operator 
represents a choice of one corner from each B-simplex, it is not hard 
to see what the requirement is. The $N^2$ point face operators must 
be chosen so that for any pair of them, the choice of corner must be the same 
for one B-simplex, and different for all the others. This is a combinatorial 
problem---and the solution of this combinatorial problem will turn out 
to be an affine plane.

To begin, let us make choices for the first two B-simplices. Since there 
are only $N^2$ choices we must use all of them, and we use 
these choices to organize a square array of $N^2$ points. Using 
$N = 3$ as our example, the result of this preliminary step is   

\begin{equation} \begin{array}{ccccc} \triangle \hspace{-4mm} {\ }^\bullet 
\hspace{3.2mm} \triangle \hspace{-2.6mm} {\ }_{\bullet} \hspace{2mm} 
\triangle \ \triangle & \ 
& \triangle \hspace{-5mm} {\ }_{\bullet} \hspace{4mm} 
\triangle \hspace{-2.6mm} {\ }_{\bullet} \hspace{2mm} \triangle \ \triangle & 
\ & \triangle \hspace{-2.6mm} {\ }_{\bullet} \hspace{2mm} 
\triangle \hspace{-2.6mm} {\ }_{\bullet} \hspace{2mm} \triangle \ \triangle \\
\ & \ & \ & \ & \ \\
\triangle \hspace{-4mm} {\ }^\bullet 
\hspace{3.2mm} \triangle \hspace{-5mm} {\ }_{\bullet} \hspace{4.2mm} 
\triangle \ \triangle 
& \ & \triangle \hspace{-5mm} {\ }_{\bullet} 
\hspace{4.2mm} \triangle \hspace{-5mm} {\ }_{\bullet} \hspace{4.2mm} 
\triangle \ \triangle & \ &  
\triangle \hspace{-2.6mm} {\ }_{\bullet} \hspace{2mm} 
\triangle \hspace{-5mm} {\ }_{\bullet} \hspace{4.2mm} 
\triangle \ \triangle \\ 
\ & \ & \ & \ & \ \\
\triangle \hspace{-4mm} {\ }^\bullet 
\hspace{3.2mm} \triangle \hspace{-4mm} {\ }^\bullet 
\hspace{3.2mm} \triangle \ \triangle 
& \ & \triangle \hspace{-5mm} {\ }_{\bullet} \hspace{4mm} 
\triangle \hspace{-4mm} {\ }^\bullet 
\hspace{3.2mm} \triangle \ \triangle & \ & 
\triangle \hspace{-2.6mm} {\ }_{\bullet} \hspace{2mm} 
\triangle \hspace{-4mm} {\ }^\bullet 
\hspace{3.2mm} \triangle \ \triangle \end{array} 
\end{equation}

\

\noindent The understanding is that we must make some definite choice also 
for the $N-1$ B-simplices that have been left empty here. Let us consider the 
first of these. Interestingly, the condition that we extracted from eq. (\ref{17}) 
means that we are looking for a Latin square: The problem is to place letters 
from an alphabet of $N$ letters on a square array, in such a way that a given 
letter occurs exactly once in each row and exactly once in each column. 
There are many Latin squares. The difficulties begin when we come to the 
next step, since there are now two Latin squares that force consistency 
conditions on each other. Mathematically, we are facing the problem of finding 
altogether $N-1$ Latin squares that are mutually orthogonal: By definition 
a pair of Latin squares is said to be orthogonal if every pair of letters, one 
from each Latin square, defines a unique position in the array \cite{Bennett}. 

For the $N = 3$ case, it is enough to find one pair of orthogonal Latin squares, and 
by referring back to the picture of the pencils of parallel lines in the affine 
plane of order $N$, we realize that the third and fourth pencil define such 
a pair:

\begin{equation} \mbox{3d pencil:} \hspace{4mm} 
\begin{array}{ccc} \triangle \hspace{-5mm} {\ }_{\bullet} \hspace{4mm} 
& \triangle \hspace{-2.6mm} {\ }_{\bullet} \hspace{2mm} & 
\triangle \hspace{-4mm} {\ }^\bullet 
\hspace{3.2mm}\\
\ & \ & \ \\
\triangle \hspace{-2.6mm} {\ }_{\bullet} \hspace{2mm} & 
\triangle \hspace{-4mm} {\ }^\bullet 
\hspace{3.2mm}  & \triangle \hspace{-5mm} {\ }_{\bullet} \hspace{4mm} \\ 
\ & \ & \ \\
\triangle \hspace{-4mm} {\ }^\bullet 
\hspace{3.2mm} 
& \triangle \hspace{-5mm} {\ }_{\bullet} \hspace{4mm} & 
\triangle \hspace{-2.6mm} {\ }_{\bullet} \hspace{2mm} 
\end{array} \hspace{8mm} 
\hspace{8mm} \mbox{4th pencil:} \hspace{4mm} 
\begin{array}{ccc} 
\triangle \hspace{-2.6mm} {\ }_{\bullet} \hspace{2mm}
& \triangle \hspace{-4mm} {\ }^\bullet 
\hspace{3.2mm} & 
\triangle \hspace{-5mm} {\ }_{\bullet} \hspace{4mm} \\
\ & \ & \ \\
\triangle \hspace{-5mm} {\ }_{\bullet} \hspace{4mm} & 
\triangle \hspace{-2.6mm} {\ }_{\bullet} \hspace{2mm} & 
\triangle \hspace{-4mm} {\ }^\bullet 
\hspace{3.2mm} \\ 
\ & \ & \ \\
\triangle \hspace{-4mm} {\ }^\bullet 
\hspace{3.2mm} 
& \triangle \hspace{-5mm} {\ }_{\bullet} \hspace{4mm} & 
\triangle \hspace{-2.6mm} {\ }_{\bullet} \hspace{2mm} 
\end{array} 
\end{equation}

\noindent This is not an accident. The $N+1$ pencils of parallel lines 
in an affine plane of order $N$ always define $N-1$ 
mutually orthogonal Latin squares---the first two pencils 
are used to organize the array. The converse is also true: An affine plane 
of order $N$ exists if and only if one can find $N-1$ mutually orthogonal 
Latin squares \cite{Bennett}. 

Interestingly, for $N = 6$ one cannot find even one pair of orthogonal 
Latin squares. This was conjectured by Euler and proved 120 years later 
by Tarry \cite{Tarry}.  

To continue our $N = 3$ example, it is clear that our two Latin 
squares can be used to complete the array that we began above, thus:

\begin{equation} \begin{array}{ccccc} \triangle \hspace{-4mm} {\ }^\bullet 
\hspace{3.2mm} \triangle \hspace{-2.6mm} {\ }_{\bullet} \hspace{2mm} 
\triangle \hspace{-5mm} {\ }_{\bullet} \hspace{4mm} 
\triangle \hspace{-2.6mm} {\ }_{\bullet} \hspace{2mm} & \ 
& \triangle \hspace{-5mm} {\ }_{\bullet} \hspace{4mm} 
\triangle \hspace{-2.6mm} {\ }_{\bullet} \hspace{2mm} 
\triangle \hspace{-2.6mm} {\ }_{\bullet} \hspace{2mm} 
\triangle \hspace{-4mm} {\ }^\bullet \hspace{3.2mm} & 
\ & \triangle \hspace{-2.6mm} {\ }_{\bullet} \hspace{2mm} 
\triangle \hspace{-2.6mm} {\ }_{\bullet} \hspace{2mm} 
\triangle \hspace{-4mm} {\ }^\bullet \hspace{3.2mm} 
\triangle \hspace{-5mm} {\ }_{\bullet} \hspace{4.2mm} \\
\ & \ & \ & \ & \ \\
\triangle \hspace{-4mm} {\ }^\bullet 
\hspace{3.2mm} \triangle \hspace{-5mm} {\ }_{\bullet} \hspace{4.2mm} 
\triangle \hspace{-2.6mm} {\ }_{\bullet} \hspace{2mm} 
\triangle \hspace{-5mm} {\ }_{\bullet} \hspace{4.2mm} 
& \ & \triangle \hspace{-5mm} {\ }_{\bullet} 
\hspace{4.2mm} \triangle \hspace{-5mm} {\ }_{\bullet} \hspace{4.2mm} 
\triangle \hspace{-4mm} {\ }^\bullet 
\hspace{3.2mm} 
\triangle \hspace{-2.6mm} {\ }_{\bullet} \hspace{2mm} & \ &  
\triangle \hspace{-2.6mm} {\ }_{\bullet} \hspace{2mm} 
\triangle \hspace{-5mm} {\ }_{\bullet} \hspace{4.2mm} 
\triangle \hspace{-5mm} {\ }_{\bullet} \hspace{4mm} 
\triangle \hspace{-4mm} {\ }^\bullet \hspace{3.2mm} \\ 
\ & \ & \ & \ & \ \\
\triangle \hspace{-4mm} {\ }^\bullet 
\hspace{3.2mm} \triangle \hspace{-4mm} {\ }^\bullet 
\hspace{3.2mm} \triangle \hspace{-4mm} {\ }^\bullet \hspace{3.2mm} 
\triangle \hspace{-4mm} {\ }^\bullet \hspace{3.2mm} 
& \ & \triangle \hspace{-5mm} {\ }_{\bullet} \hspace{4mm} 
\triangle \hspace{-4mm} {\ }^\bullet \hspace{3.2mm} 
\triangle \hspace{-5mm} {\ }_{\bullet} \hspace{4mm} 
\triangle \hspace{-5mm} {\ }_{\bullet} \hspace{4mm} & \ & 
\triangle \hspace{-2.6mm} {\ }_{\bullet} \hspace{2mm} 
\triangle \hspace{-4mm} {\ }^\bullet \hspace{3.2mm} 
\triangle \hspace{-2.6mm} {\ }_{\bullet} \hspace{2mm}  
\triangle \hspace{-2.6mm} {\ }_{\bullet} \hspace{2mm} 
\hspace{3.2mm} \end{array} 
\end{equation}

\

\noindent This selection of point face operators tells us how to 
inscribe the D-simplex in the Complementarity Polytope. 

The procedure works if and only if an affine 
plane of order $N$ exists. Any affine plane of order $N$ will 
do---we are not restricted to those that can be coordinatized with finite 
fields. If $N$ is such that no affine plane of order $N$ exists, the 
D-simplex cannot be inscribed in the Complementarity Polytope. For values of 
$N$ such that an affine plane exists, we can proceed to reproduce all the key 
formulas in the paper by Gibbons, Hoffman and Wootters \cite{Wootters3}. 
First we reindex our selection of point face operators as $A_{ij}$, where 
$1 \leq i,j \leq N$. The matrix that represents a corner of the 
Complementarity Polytope can be recovered by summing all the $A_{ij}$ 
that sit on the line representing that corner:

\begin{equation} P_{\omega} = \frac{1}{N}\sum_{line \ \omega}A_{ij} \ . 
\end{equation}

\noindent We can also use the fact that our $A_{ij}$ form a basis in the 
space of matrices to define Wootters' Wigner function $W_{ij}$, which is 
indeed a function of position in the affine plane:

\begin{equation} W_{ij} = \frac{1}{N}\mbox{Tr}A_{ij}\rho \ . \end{equation}

\noindent This equation can easily be solved for $\rho$ in terms of $W_{ij}$. 
Given a Wigner function, we can assign a set of numbers 
$p_{\omega}$ to the lines in such a way that $\sum p_{\omega} = 1$, where 
the sum goes over all lines belonging to a given pencil of parallel 
lines. We do this 
by summing the Wigner function along the line: 

\begin{equation} p_{\omega} \equiv \sum_{line \ \omega}W_{ij} = 
\mbox{Tr}P_{\omega}\rho \ . \label{p} \end{equation}

\noindent And so on \cite{Wootters3}. 

Make sure to notice that all of this is true, regardless of the existence of 
the complete set of MUBs. All that is needed is the Complementarity Polytope 
as such, together with an inscribed $N^2-1$ dimensional D-simplex. 

\vspace{1cm}

{\bf 5. MUBs and conclusions}

\vspace{5mm}
 
\noindent Seemingly, we have arrived at all the conclusions of Gibbons 
et al. \cite{Wootters3}, without serious work. In fact we have not, 
because we have tried to settle the account without consulting the host. 
Although the Wigner formulation seems to work in happy obliviousness of 
the question whether the corners are quantum states or not, there is a 
catch: For the probability interpretation to work, we need $p_{\omega} 
\geq 0$ in equation (\ref{p}). But the only way to ensure this, for all 
positive operators $\rho$, 
is to insist that the $P_{\omega}$ have non-negative spectrum. 
Therefore we must ask whether it is possible to rotate the Complementarity 
Polytope in such a way that it defines a subset of the convex body of 
density matrices. This is our reformulation of the standard problem of 
defining a complete set of MUBs, but it leads to no new insights into it. 
All I can do is to refer you to the literature, where various mathematical 
approaches are employed in clever ways \cite{Anton1, Vatan, Rubin, 
Klappenecker,Durt} that so far have borne fruit only for $N = p^k$. For 
$N = 6$, Zauner's suspicion is that only 3 MUBs exist \cite{Zauner}, but 
nobody knows for sure. Wootters has recently reviewed the whole 
question \cite{Wootters4}.

Incidentally, for any polytope in $N^2-1$ dimensions, 
we can ask if it can be oriented so that its corners are pure quantum 
states. If this can be done for a regular D-simplex the result is called 
a Symmetric Informationally Complete POVM \cite{Renes}. It would be nice 
to report that the D-simplex that we inscribed in the Complementarity 
Polytope defines a MUB-SIC-POVM---to pick an acronym---when the corners 
of the original polytope become 
MUBs. Unfortunately, at least if we make use of the arguably most elegant 
choice of face point operators \cite{Wootters1}, this is so only for 
$N = 2$ and $N = 3$.   

In conclusion, in the language of the Complementarity Polytope, an affine 
plane of order $N$ exists if and only if it is possible to 
inscribe a regular simplex in the Complementarity Polytope, with the $N^2$ 
corners of the former situated in the middle of the facets of the latter. 
The problem of finding $N+1$ MUBs is equivalent to 
the problem of rotating the Complementarity Polytope so that its corners 
sit inside the convex body of density matrices. On the face of it, these 
problems are very different. It 
appears that the finite geometries are related to MUBs much like 
the icing on a cake---it makes the cake more delicious, but it has 
nothing to do with the existence of the cake. This may be a superficial 
view, however. Only the Baker knows. 

\vspace{1cm}

{\bf Acknowledgement:}

\vspace{5mm}

\noindent I thank Gunnar Bj\"ork and Bengt Nagel for discussions, and 
\AA sa Ericsson and Ninos Poli for many 
discussions---and of course Andrei Khrennikov for inviting me to visit 
Sm\aa land.

\end{document}